\documentclass[runningheads]{llncs}

\usepackage[T1]{fontenc}
\usepackage{graphicx}
\usepackage{url}
\usepackage{cite}

\usepackage{amsmath}
\usepackage{subcaption}
\usepackage{listings}

\usepackage{xcolor}

\definecolor{CodePurple}{RGB}{139,00,139}

\definecolor{pblue}{rgb}{0.13,0.13,1}
\definecolor{pgreen}{rgb}{0,0.5,0}
\definecolor{pred}{rgb}{0.9,0,0}
\definecolor{pgrey}{rgb}{0.46,0.45,0.48}
\newcommand{\codeinline}[1]{\lstinline[basicstyle=\ttfamily, keywordstyle=, language=Java]|#1|}

\begin{document}
\title{Coverage Isn’t Enough: SBFL-Driven Insights into Manually Created vs. Automatically Generated Tests}
%
%
\author{Sasara Shimizu\inst{1} \and
Yoshiki Higo\inst{1}}
\authorrunning{S. Shimizu et al.}
%
\institute{The University of Osaka, Osaka, Japan \\
\email{\{simizu-s,higo\}@ist.osaka-u.ac.jp}}

\maketitle

\begin{abstract}
The testing phase is an essential part of software development, but manually creating test cases can be time-consuming. Consequently, there is a growing need for more efficient testing methods. To reduce the burden on developers, various automated test generation tools have been developed, and several studies have been conducted to evaluate the effectiveness of the tests they produce.
However, most of these studies focus primarily on coverage metrics, and only a few examine how well the tests support fault localization—particularly using artificial faults introduced through mutation testing.
In this study, we compare the SBFL (Spectrum-Based Fault Localization) score and code coverage of automatically generated tests with those of manually created tests. The SBFL score indicates how accurately faults can be localized using SBFL techniques. By employing SBFL score as an evaluation metric—an approach rarely used in prior studies on test generation—we aim to provide new insights into the respective strengths and weaknesses of manually created and automatically generated tests.
Our experimental results show that automatically generated tests achieve higher branch coverage than manually created tests, but their SBFL score is lower, especially for code with deeply nested structures. These findings offer guidance on how to effectively combine automatically generated and manually created testing approaches.

\end{abstract}

\section{Introduction}

In software development, unit testing is essential for improving code quality and detecting bugs early. However, manually creating unit tests is time-consuming and challenging, especially under tight deadlines and limited human resources. In large-scale projects, achieving comprehensive test coverage is often impractical.

To address these challenges, various automated unit test generation tools have been developed. These tools analyze program code to automatically generate test cases, thereby reducing the burden on developers. Notable examples include EvoSuite~\cite{EvoSuite}, Randoop~\cite{randoop}, and Agitar~\cite{agitar}.

Recent advances in test generation have shown that combining different tools can improve fault detection~\cite{fraser}. Several studies have evaluated such tools from multiple perspectives, including code coverage, mutation score, and bug detection~\cite{serra2019msr, jtexpert}. 

While automatically generated tests perform well in terms of coverage and mutation analysis, they often struggle to identify real faults. This has led to the suggestion that combining manually created and automatically generated tests yields better outcomes. For example, EvoSuite\textsubscript{Amp}~\cite{Roslan2022}, which leverages developer-written tests as seeds, has demonstrated improved fault detection, albeit at the cost of readability. Other studies have identified the limitations of test generation tools, such as challenges in object construction, large search spaces, and handling multithreaded code~\cite{Herlim2021, watanabe2023profes}.

Spectrum-Based Fault Localization (SBFL) is a technique used to identify defect locations within a program. The SBFL score indicates how accurately faults can be localized using SBFL techniques. In the following sections, we refer to manually created tests as MC-tests and automatically generated tests as AG-tests.

While AG-tests have been widely studied, few works directly compare them with MC-tests using real-world programs. Moreover, their effectiveness has rarely been evaluated from the perspective of fault localization. In particular, no prior study has investigated the use of SBFL score derived from mutation-based artificial faults to compare these test types.

In this study, we compare MC-tests and AG-tests from two perspectives: SBFL score and code coverage. Our objective is to analyze the performance of each test type to identify their respective strengths and weaknesses. Through this comparison, we aim to provide insights into how MC-tests and AG-tests can be effectively combined and under what circumstances each is most appropriate.

We address the following research questions:  
\textbf{RQ1:}Which test type achieves better code coverage—AG-tests or MC-tests?  
\textbf{RQ2:} Which test type performs better in terms of SBFL score?

The results of RQ1 indicate that AG-tests are effective for testing simple conditional branches and structural patterns. They are particularly well suited for achieving broad coverage with minimal manual effort.

However, RQ2 reveals that high coverage does not necessarily imply effective fault localization. In particular, for code with deeply nested logic, the SBFL performance of AG-tests tends to degrade, and they are often outperformed by MC-tests. This suggests that simply maximizing coverage is insufficient. When code complexity increases, SBFL score tends to decline, limiting the utility of AG-tests.

These findings suggest that combining MC-tests with AG-tests can leverage the strengths of both, enabling the construction of more effective and comprehensive test suites.

\section{Background}

Herein, we describe the background and the research objectives of this study.

\subsection{Unit Test}

In software development, software testing is conducted to verify whether a program functions as intended. Software testing is performed at multiple stages, depending on the level of testing. Unit testing is performed on the smallest unit of code, such as functions or methods. Since this process is performed early in the development cycle, it helps to identify bugs and issues at an early stage. As a result, unit testing has become an essential practice in software development.

To improve the efficiency of unit testing, test automation frameworks are widely used. A test automation framework provides the necessary environment and utilities for writing and executing test cases. It offers features such as test execution automation and result reporting. For example, JUnit is a widely used test automation framework for Java and is supported by integrated development environments (IDEs) such as Eclipse and IntelliJ IDEA.

To conduct unit testing, a test suite corresponding to the target code must be prepared. A test suite is a collection of test cases designed to achieve specific testing objectives. On the other hand, a test case is the smallest unit of testing and consists of specific inputs to the target code and the expected results.

\subsection{Automated Test Generation}

Manually creating a unit test suite requires a significant amount of effort. To address this challenge, research has been conducted on automatic unit test generation. Notable tools in this field include EvoSuite and Randoop. In this study, we utilize EvoSuite~\cite{EvoSuite}.

EvoSuite is a tool designed to automatically generate unit tests for Java projects. It generates JUnit-format test suites for individual Java classes, leveraging exploratory approaches such as hybrid search, dynamic symbolic execution, and testability transformation to maximize code coverage. Initially, multiple random test cases are generated, which are then iteratively refined through exploratory approaches. The resulting test suite is minimized while maintaining coverage criteria, ensuring that the generated unit tests remain as concise as possible. EvoSuite has been widely used in existing research.

\subsection{Research Objective}

In this study, we compare tests created by developers with those generated by automated tools from two perspectives: SBFL (Spectrum-Based Fault Localization) score and code coverage. Our objective is to analyze the performance of manually created and automatically generated test cases to identify their respective strengths and weaknesses. By employing the SBFL score, which has rarely been addressed in previous research, we expect to reveal new insights into the characteristics of each testing approach.

Through this comparison, we aim to provide insights into how to effectively combine MC-tests and AG-tests and determine the appropriate scenarios for their use. Ultimately, this research is expected to help improve the test design process in software development environments.

\section{Experiment Setup}

Herein, we describe the research questions, the evaluation metrics, the target programs, and the experimental procedure used in this study.

\subsection{Research Questions}

In this experiment, we address the following research questions.

\begin{itemize}
    \item[\textbf{RQ1}] Which tests perform better in code coverage between AG-tests and MC-tests?
    \item[\textbf{RQ2}] Which tests perform better in  SBFL score between AG-tests and MC-tests?
\end{itemize}

\subsection{Evaluation Metrics}
This section explains the code coverage and SBFL score used as evaluation metrics.

\subsubsection{Code coverage}

Code coverage is a metric that indicates the proportion of code exercised by tests. Coverage criteria define the methods used to measure this coverage, with statement coverage and branch coverage being among the most widely adopted. In this study, we employ both statement and branch coverage, as described below.

Statement Coverage: This metric measures the proportion of executable lines of code in the target code that are executed by the test suite. When calculating code coverage using statement coverage, the following formula is used:

\begin{equation}
\begin{split}
    \text{Statement Coverage} = \frac{\text{Number of lines executed by the test suite}}{\text{Number of executable lines in the target code}}
\end{split}
\end{equation}

Branch coverage measures whether the test suite covers both the true and false outcomes of branch conditions within the target code. When calculating code coverage using branch coverage, the following formula is used:

\begin{equation}
\begin{split}
    \text{Branch Coverage} = \frac{\text{Number of branches executed by the test suite}}{\text{Number of branches in the target code}}
\end{split}
\end{equation}

\subsubsection{SBFL}

One technique for estimating defect locations in a program is Spectrum-Based Fault Localization (SBFL)~\cite{sbfl}. 
This technique makes use of execution path information recorded during test execution.
The underlying idea is that statements executed in failed test cases are more likely to contain defects, while statements executed in successful test cases are less likely to be defective.

First, all tests are executed, and both the test results (pass/fail) and execution path information are recorded. Based on this information, a suspiciousness score, which indicates the likelihood of a statement containing a defect, is calculated for each statement. The suspiciousness score, $susp(s)$, is calculated using the Ochiai~\cite{ochiai} formula, as shown in Equation~\ref{eq:susp}:

\begin{equation}
susp(s) = \frac{fail(s)}{\sqrt{totalFail \times \left(fail(s) + pass(s)\right)}}
\label{eq:susp}
\end{equation}

\begin{itemize}
    \item $fail(s)$: the number of failed test cases that executed statement $s$
    \item $pass(s)$: the number of successful test cases that executed statement $s$
    \item $totalFail$: the total number of failed test cases
\end{itemize}

The suspiciousness score $susp(s)$ is computed for all statements, and statements with higher scores are more likely to contain defects.

\subsubsection{SBFL Score}

Sasaki et al. proposed SBFL Suitability~\cite{sasaki_en}, a metric that indicates how suitable a program is for Spectrum-Based Fault Localization (SBFL). The SBFL score is defined as the numerical value representing this suitability, ranging from 0 to 1, where higher values indicate greater suitability for SBFL.

\begin{figure}
    \centering
    \includegraphics[width=0.8\linewidth]{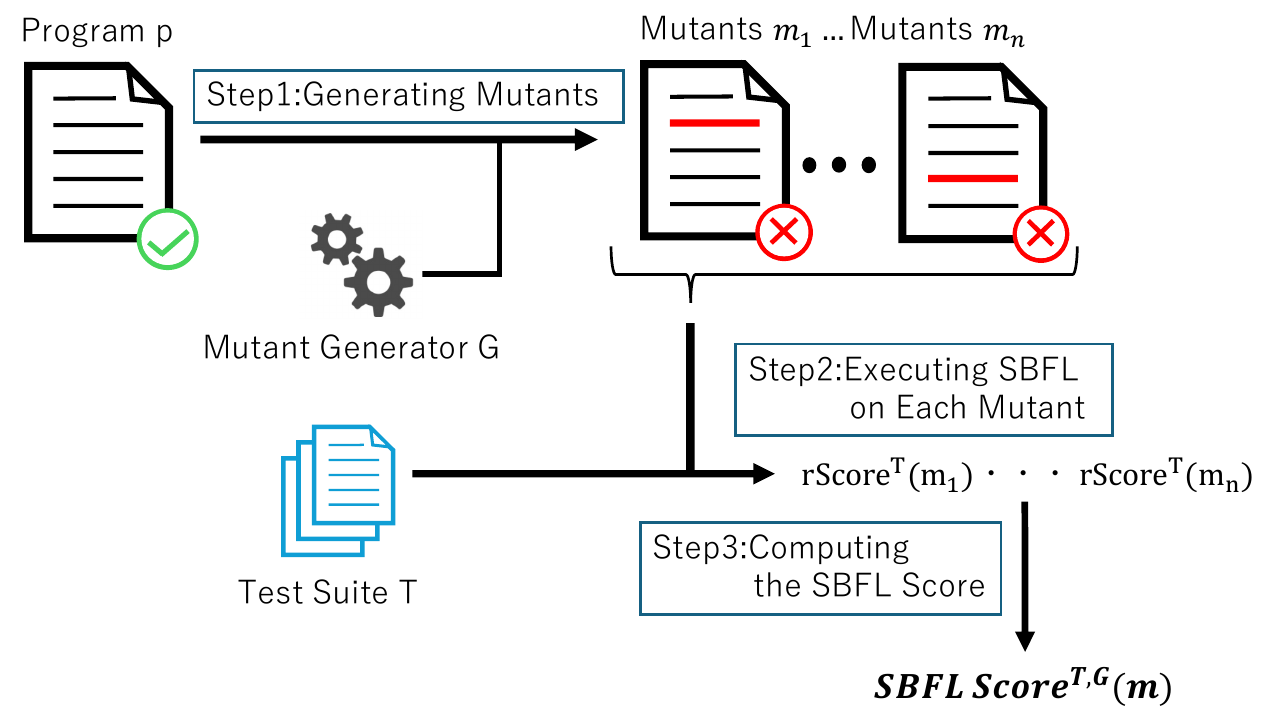}
    \caption{How to get SBFL Score}
    \label{fig:howtogetSBFLScore}
\end{figure}

The SBFL score for a program p depends on the test suite $T$ and the mutant generator $G$, and is denoted as $SBFLScore^{T,G}(p)$. Figure~\ref{fig:howtogetSBFLScore} illustrates the process of calculating the SBFL score, which consists of the following three steps:

\begin{itemize}
    \item[\textbf{Step1}] Generating mutants from the program $p$
    \item[\textbf{Step2}] Executing SBFL on each mutant and ranking suspiciousness values of the artificial faults
    \item[\textbf{Step3}] Computing the SBFL score based on the rankings of the artificial faults in each mutant 
\end{itemize}

\textbf{Step1. Generating mutants}

A target program and its test suite are prepared. Mutants are generated using a mutant generator, with each mutant differing from the original program by only a single modification. Since there can be multiple modifiable locations in a program, multiple mutants are generated. The set of mutants generated by $G$ from $P$ is denoted as $M^G(P)$.

\textbf{Step2. Executing SBFL on mutants}

SBFL is executed on each mutant in $M^G(P)$ using the test suite $T$, and the suspiciousness values for each statement are calculated. For a given mutant $m\in M^G(P)$, the following are defined:

\begin{itemize}
    \item $susp^T(s)$ : Suspiciousness value of statement $s$
    \item $rank^T(s)$ : Rank of the suspiciousness value for statement $s$
    \item  $rScore^T(s)$ : Normalized rank of the suspiciousness value for statement s
\end{itemize}

The Ochiai formula is used to calculate suspiciousness values.
We chose the Ochiai formula because it is one of the most widely used metrics in the field of SBFL. Many prior studies have demonstrated its effectiveness, and it has become a de facto standard in evaluating suspiciousness~\cite{Sarhan2022}. Therefore, in this study, we also adopt the Ochiai formula as a standard approach to ensure comparability with existing work.

The ranking of suspiciousness values is determined by sorting statements in descending order of suspiciousness. If two statements have the same highest suspiciousness value (e.g., two statements with $sus p$ = 1.0 ), they are both ranked second, and the next highest (e.g., $susp$ = 0.8 ) is ranked third.

Since the meaning of rank values depends on the total number of statements, ranks are normalized within the range of 0 to 1. For example, ranking 10th out of 10 statements differs from ranking 10th out of 100 statements, with the latter having a higher significance. The normalized rank $rScore^T(s)$ of a statement $s$ is computed as follows:

\begin{equation}
    rScore^T(s) = 1 - \frac{rank^T(s) - 1}{totalStatements^T - 1}
    \label{eq:rscore}
\end{equation}

where $totalStatements^T$ represents the total number of statements executed by test suite $T$. A value of 1 represents the highest rank (most useful), while 0 represents the lowest.

The normalized suspiciousness rank for a mutant $m$, denoted as $rScore^T(m)$, is defined as the normalized rank of the artificial fault statement $s^m_{fault}$ within the mutant:

\begin{equation}
    rScore^T(m) = rScore^T(s^m_{fault})
\end{equation}

\textbf{Step3. Computing the SBFL score}

The SBFL score is computed as the average $rScore$ over all generated mutants in $\left| M^G(P) \right|$. The total number of mutants is denoted as $\left| M^G(P) \right|$, and the final SBFL score is calculated as:

\begin{equation}
    SBFLScore^{T,G}(P) = \frac{1}{|M^G(P)|} \sum_{m \in M} rScore^T(m)
\end{equation}

A higher SBFL score indicates that the program is well-suited for fault localization using SBFL.

\subsection{Experimental Targets}

In this experiment, we measured statement and branch coverage as well as SBFL score for tests written in Java, and compared the test cases.

We used Defects4J (version 3.0.1)~\cite{defects4j} for this experiment, a dataset that collects real-world bugs from Java projects. Defects4J contains buggy code, fixed code, and developer-created tests. 
We measured coverage and SBFL score on 167 datasets from the Lang and Math projects in Defects4J, which have few dependencies and can be compiled relatively easily.

\subsection{Experimental Procedure}

In this experiment, we compare MC-tests, which are the developer-written tests included in Defects4J, with AG-tests produced by EvoSuite (version 1.2.0). For each bug, AG-tests were generated for the fixed version of the class where the bug originally existed. Specifically, EvoSuite was run on the fixed code to ensure that generated tests could exercise the corrected behavior and reach all relevant code paths. The target classes were therefore the ones that contained bugs and had been fixed in Defects4J; these are the same classes used during test generation with EvoSuite.

We acknowledge that this procedure differs from the process developers follow when writing tests manually, as developers typically write tests against buggy or evolving code rather than a fully fixed version. Our choice to use the fixed version was intended to ensure that EvoSuite-generated tests would fully exercise the intended functionality and allow a fair evaluation of coverage and SBFL metrics. While this setup may not exactly mimic the original developer workflow, it provides a consistent basis for comparing AG-tests with MC-tests in terms of their coverage and fault localization effectiveness.

To evaluate the tests, we measured both code coverage and SBFL score. For coverage, we used JaCoCo\footnote{https://www.jacoco.org/jacoco/}, focusing on the class files where the bugs occurred and were fixed—that is, the same class files targeted by EvoSuite. We integrated JaCoCo using the Ant task framework provided by Defects4J. JaCoCo calculates both statement and branch coverage.

\begin{table}[h]
\centering
\caption{Mutation operators}
\begin{tabular}{lcc} \hline
Mutation operators & Original conditional & Mutated conditional \\ \hline
Conditionals Boundary & \codeinline{a<b} & \codeinline{a<=b} \\
Increments & \codeinline{n++} & \codeinline{n--} \\
Invert Negatives & \codeinline{-n} & \codeinline{n} \\
Math & \codeinline{a+b} & \codeinline{a-b} \\
Negate Conditionals & \codeinline{a==b} & \codeinline{a!=b} \\
Void Method Calls & \codeinline{method();} & \codeinline{;} \\
Primitive Returns & \codeinline{return 5;} & \codeinline{return 0;} \\ \hline
\label{table:mutation_op}
\end{tabular}
\end{table}

 For mutation-based fault localization, we followed the procedure described in Section 3.2, Steps 1 through 3. Mutants were generated using Mutanerator\footnote{https://github.com/kusumotolab/Mutanerator}, targeting the same fixed class files. The mutation operators used are listed in Table~\ref{table:mutation_op}. To obtain the information necessary for computing the SBFL score, we employed Gzoltar\footnote{https://github.com/GZoltar/gzoltar}.

It should be clarified that Gzoltar itself is not specifically designed to directly provide the final SBFL score for our evaluation. Instead, Gzoltar is a tool for test execution and for calculating suspiciousness values of program elements. In our study, we used Gzoltar to obtain these suspiciousness values and ranking information, and then used this intermediate data to derive the SBFL score required for our evaluation.

For the MC-tests, the number of test cases ranged from 3 to 2,447, with a mean of 102.58. 
For the AG-tests, the number of test cases ranged from 4 to 642, with a mean of 112.14. 
Although the maximum size of the MC-tests was considerably larger due to a small number of outlier projects, the overall averages of the two types of test suites are comparable. 

As for the mutation analysis, a total of 13,444 mutants were generated in our study. 
The mutation operators used are listed in Table~\ref{table:mutation_op}.

\subsubsection{Analysis Approach for RQ1}

In RQ1, we measure coverage and visualize the overall results using box plots to compare AG-tests and MC-tests. Furthermore, we conduct the Wilcoxon signed-rank test to examine whether there is a statistically significant difference between the two types of tests. We also analyze specific examples of source code and test cases to better understand the differences between AG-tests and MC-tests.

\subsubsection{Analysis Approach for RQ2}

In RQ2, we perform the Wilcoxon signed-rank test on the overall results to investigate whether there is a statistically significant difference between the two types of tests. Additionally, we analyze the results for each generated mutant based on the nesting depth of the code. This includes visualization using box plots and calculation of the effect size $r$ from the Wilcoxon test. Furthermore, we analyze the results by mutation operator and visualize the distributions using box plots.

\section{Experimental Result}

We compare and analyze the results of coverage and SBFL score for MC-tests and AG-tests.

\subsection{RQ1: Coverage Comparison Results}

Bugs that failed to build in Java 11 were excluded. As a result, coverage data was obtained for 133 out of 167 bugs.

\begin{figure}[htbp]
    \centering
    \begin{minipage}[b]{0.4\linewidth}
        \centering
        \includegraphics[width=\linewidth]{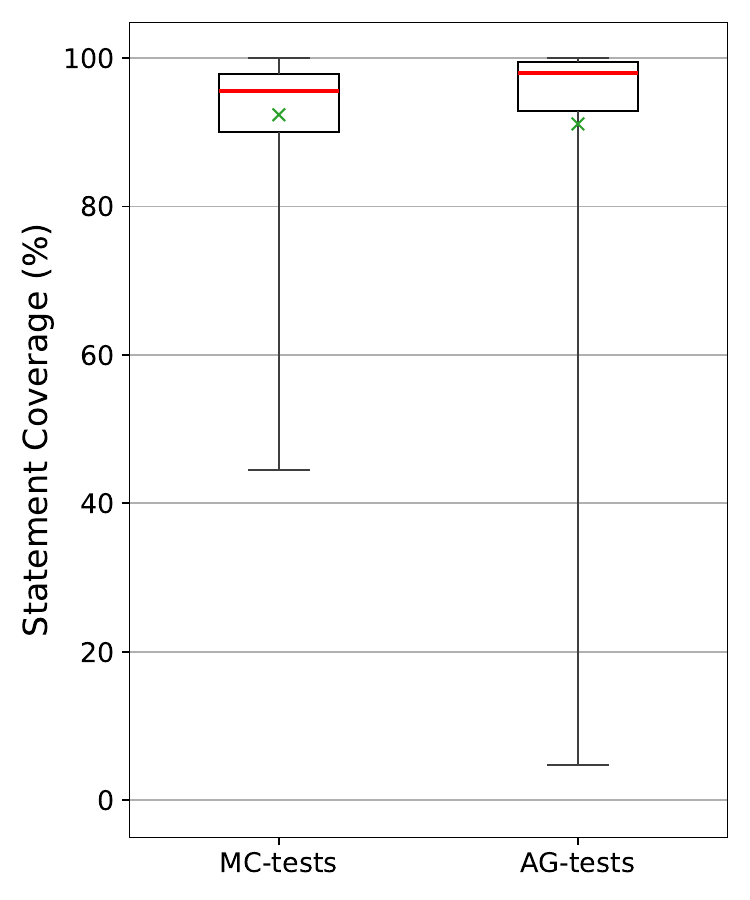}
        \caption{Statement Coverage Comparison}
        \label{fig:linecov}
    \end{minipage}
    \hspace{0.05\linewidth} 
    \begin{minipage}[b]{0.4\linewidth}
        \centering
        \includegraphics[width=\linewidth]{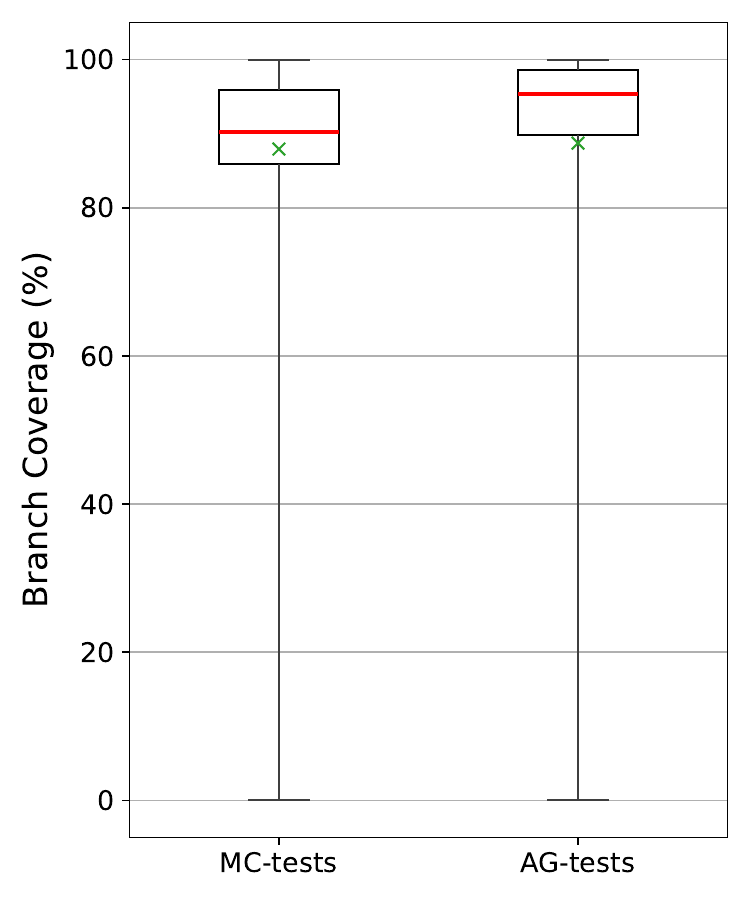}
        \caption{Branch Coverage Comparison}
        \label{fig:branchcov}
    \end{minipage}
\end{figure}

\begin{table}[b]
\caption{Comparison of average coverage}    
  \centering
  \begin{tabular}{crr}\hline
     & MC-tests& AG-tests \\ \hline
   Statement Coverage(\%) & 92.3 & 91.0 \\
   Branch Coverage(\%) & 87.9 & 88.7 \\ \hline
  \end{tabular}
  \label{table:coverage}
\end{table}

The results of statement coverage are shown in Figure~\ref{fig:linecov}, and those of branch coverage are shown in Figure~\ref{fig:branchcov}. The average coverage values are summarized in Table~\ref{table:coverage}. The red line in each box plot represents the median, while the cross mark (×) indicates the mean. For statement coverage, the range of values is wider for AG-tests than for MC-tests. In contrast, branch coverage shows no significant difference between the two. 

We conducted a Wilcoxon signed-rank test to examine whether there were statistically significant differences between MC-tests and AG-tests in terms of statement and branch coverage. The test yielded a p-value of 0.19 for statement coverage and 0.030 for branch coverage. Thus, there was no significant difference in statement coverage, while a significant difference was found in branch coverage. 

Furthermore, we conducted a one-sided Wilcoxon signed-rank test for branch coverage, setting the null hypothesis that AG-tests are not superior to MC-tests, and the alternative hypothesis that AG-tests are superior. The result yielded a p-value of 0.015, leading to the rejection of the null hypothesis at the 5\% significance level. Therefore, AG-tests significantly outperformed MC-tests in branch coverage.

To further interpret these results, the distinction between statement and branch coverage is important. The absence of a significant difference in statement coverage indicates that both MC-tests and AG-tests are generally effective at covering executable lines of code. However, the significant difference in branch coverage suggests that AG-tests have a particular advantage in covering both true and false paths of conditional statements.
Therefore, AG-tests tend to explore a wider set of branch conditions, leading to higher branch coverage even when statement coverage is comparable.

We examined the source code of bugs where MC-tests achieved higher coverage and those where AG-tests had higher coverage. Additionally, we compared the number of MC-tests and AG-tests cases.

\begin{figure}[tb]
\begin{lstlisting}[caption={Code of bugID Math-6}, label={lst:Math-6}, language=Java]
public class LevenbergMarquardtOptimizer
    extends AbstractLeastSquaresOptimizer {
    ...
    @Override
    protected PointVectorValuePair doOptimize() {
        checkParameters();
        final int nR = getTarget().length; 
        final double[] currentPoint = getStartPoint();
        final int nC = currentPoint.length; 
    ...
\end{lstlisting}
\end{figure}

In the Math project, bug ID 6 exhibits a significant difference in coverage between the MC-tests and AG-tests. The MC-tests achieve 94\% statement coverage and 86\% branch coverage. In contrast, the AG-tests achieve only 10\% statement coverage and 9\% branch coverage, resulting in a coverage difference of over 77\%. According to the source code (Code~\ref{lst:Math-6}), the issue is caused by a failure to properly set the required \texttt{OptimizationData} for the \texttt{doOptimize()} method. Although \texttt{doOptimize()} is invoked directly, the required data is missing, which results in a \texttt{NullPointerException}. Similar large differences (over 77\%) are observed for Math-38, Math-64, and Math-68, all of which involve the \texttt{doOptimize()} method. This result is consistent with the problem of object construction described in the existing study~\cite{Herlim2021}.

\textbf{Our answer to RQ1 is that AG-tests outperform MC-tests in terms of branch coverage.}

\subsection{RQ2: SBFL Score Comparison Results}

Tests that failed to build in Java 11 were excluded. As a result, SBFL score was obtained for 157 out of 167 bug IDs. 
The average SBFL score was 0.683 for manually created tests and 0.640 for generated tests.

We conducted a Wilcoxon signed-rank test to evaluate whether there was a statistically significant difference in SBFL score between MC-tests and AG-tests. The test yielded a p-value of 0.027, indicating a significant difference between the two test types.

Furthermore, we performed a one-sided Wilcoxon signed-rank test under the alternative hypothesis that MC-tests outperform AG-tests. The result yielded a p-value of 0.013, leading to the rejection of the null hypothesis at the 5\% significance level. Therefore, we conclude that MC-tests significantly outperform AG-tests in terms of SBFL score.

\begin{figure}[h]
  \begin{minipage}{0.5\columnwidth}
    \centering
    \includegraphics*[width=1\linewidth]{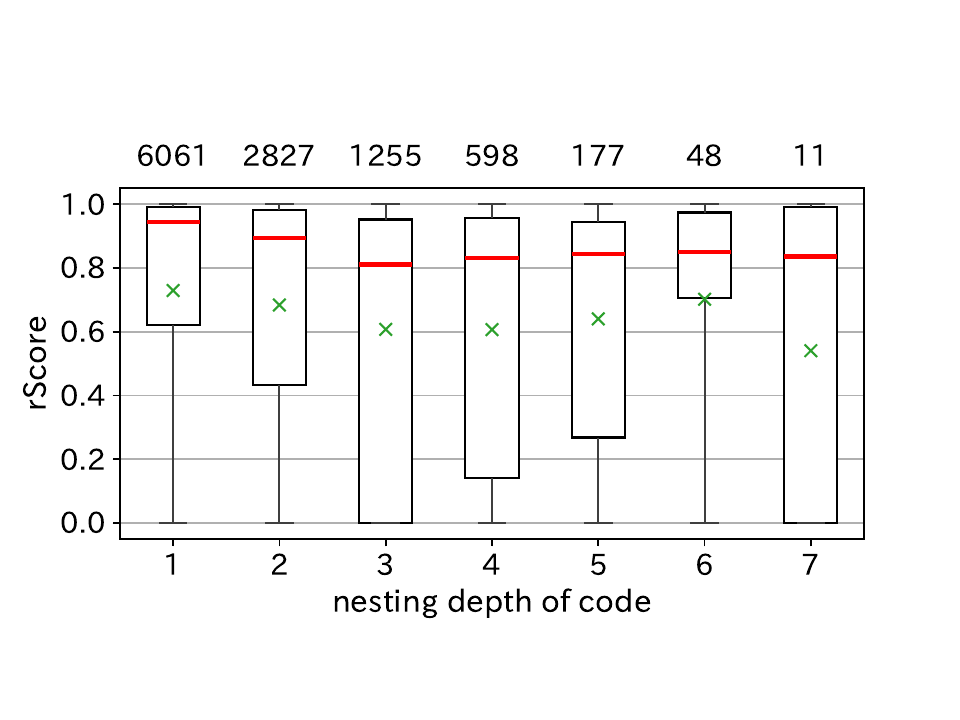}
    \subcaption{MC-tests}
    \label{fig:box_human}
  \end{minipage}
  \begin{minipage}{0.5\columnwidth}
    \centering
    \includegraphics*[width=1\linewidth]{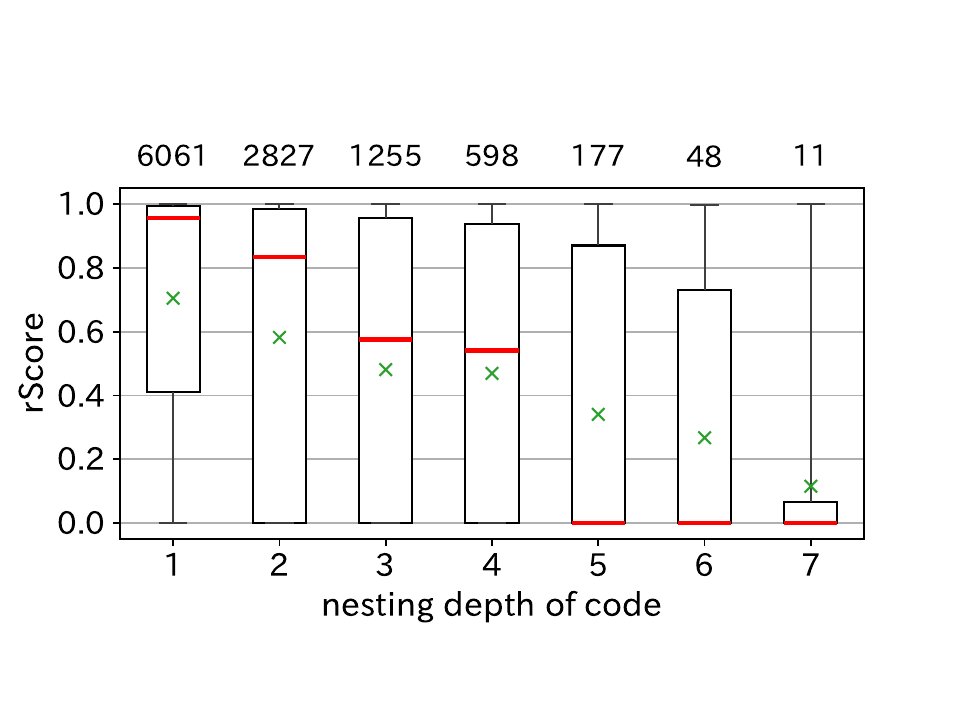}
    \subcaption{AG-tests}
    \label{fig:box_evo}
  \end{minipage}
  \caption{Relationship between $rScore$ and nesting depth of code}
\end{figure}

Figures~\ref{fig:box_human} and~\ref{fig:box_evo} show box plots illustrating the relationship between the nesting depth of mutated code and the $rScore$. The plots correspond to MC-tests and AG-tests, respectively.
The x-axis represents the nesting depth, while the y-axis shows the value of $rScore^T(s)$, calculated using Equation~\ref{eq:rscore}, which is used in computing the SBFL score. The value $rScore^T(s)$ indicates the accuracy with which the bug location in each mutant was identified. The total number of data points is 10,977, and the number of elements for each nesting depth is shown at the top of the graph.
When focusing on the medians, we observe that in MC-tests, the median remains nearly constant regardless of the nesting depth. In contrast, the median in AG-tests tends to decrease as the nesting depth increases.

\begin{table}[b]
\caption{r-value at each nesting depth}    
\centering
\begin{tabular}{cccccccc} \hline
Depth of nest & \hspace{0.5em}1\hspace{0.5em} & \hspace{0.5em}2\hspace{0.5em} & \hspace{0.5em}3\hspace{0.5em} & \hspace{0.5em}4\hspace{0.5em} & \hspace{0.5em}5\hspace{0.5em} & \hspace{0.5em}6\hspace{0.5em} & \hspace{0.5em}7\hspace{0.5em} \\ \hline
r-value       & \hspace{0.5em}$-0.095$\hspace{0.5em} & \hspace{0.5em}0.12\hspace{0.5em} & \hspace{0.5em}0.21\hspace{0.5em} & \hspace{0.5em}0.28\hspace{0.5em} & \hspace{0.5em}0.55\hspace{0.5em} & \hspace{0.5em}0.67\hspace{0.5em} & \hspace{0.5em}0.65\hspace{0.5em} \\ \hline
\end{tabular}
\label{table:r-value}
\end{table}

We conducted Wilcoxon signed-rank tests at each nesting depth to examine whether the differences in $rScore$ between MC-tests and AG-tests were statistically significant. Additionally, we calculated the effect size r-value for each nesting level, as shown in Table~\ref{table:r-value}. The effect size r-value indicates the magnitude of the observed difference, with larger absolute values representing a stronger effect.
As shown in Table~\ref{table:r-value}, the difference in $rScore$ between MC-tests and AG-tests tends to increase with greater nesting depth. This suggests that AG-tests become less effective at accurately identifying faults in deeply nested code structures.

\begin{figure}[h]
  \begin{minipage}{0.5\columnwidth}
    \centering
    \includegraphics[width=\columnwidth]{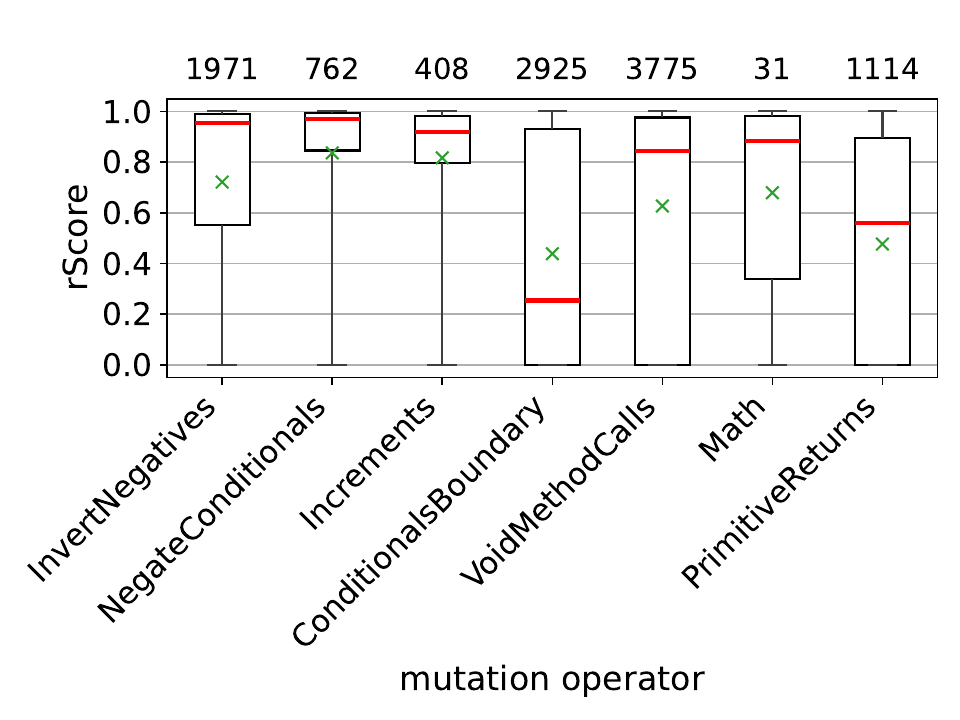}
    \subcaption{MC-tests}
    \label{fig:box_mu_human}
  \end{minipage}
  \begin{minipage}{0.5\columnwidth}
    \centering
    \includegraphics[width=\columnwidth]{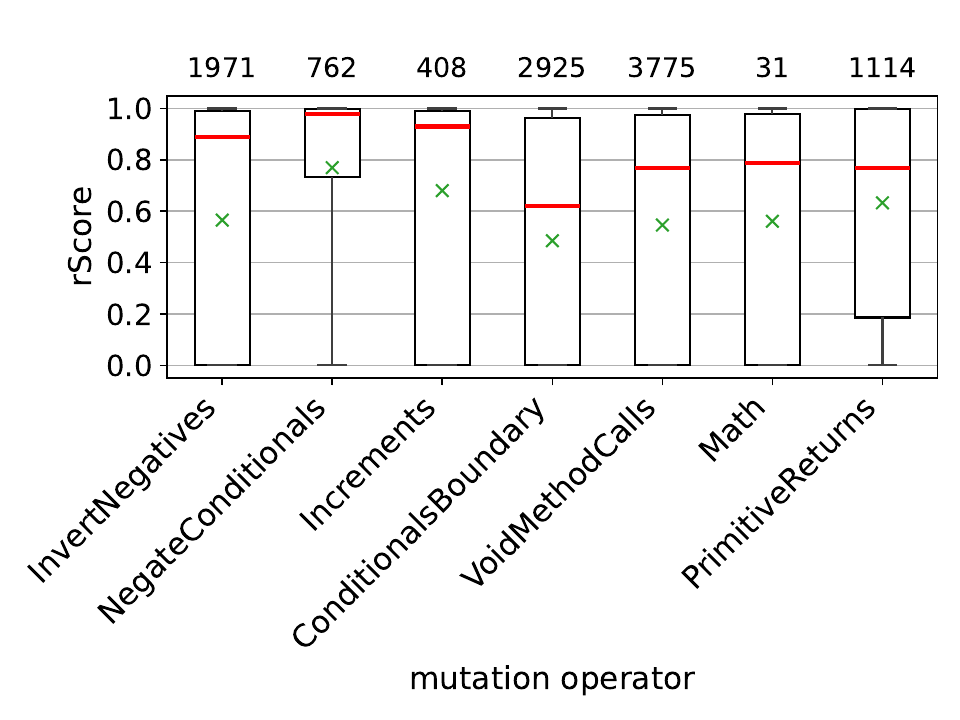}
    \subcaption{AG-tests}
    \label{fig:box_mu_evo}
  \end{minipage}
  \caption{Relationship between $rScore$ and mutation operators}
\end{figure}

Figures~\ref{fig:box_mu_human} and~\ref{fig:box_mu_evo} show box plots for MC-tests and AG-tests, respectively, illustrating the relationship between $rScore$ and the applied mutation operators.
Examples of the modifications introduced by each mutation operator are provided in Table~\ref{table:mutation_op}.
When focusing on the medians, we observed a notable difference between MC-tests and AG-tests for the ConditionalsBoundary mutation operator.

\textbf{Our answer to RQ2 is that MC-tests perform better than AG-tests in SBFL score.}

\section{Discussion}

From the results of RQ1, we found that there was no significant difference between MC-tests and AG-tests in terms of statement coverage. Moreover, AG-tests achieved higher branch coverage than MC-tests. On the other hand, the results of RQ2 revealed that AG-tests tended to yield lower SBFL score compared to MC-tests. Furthermore, the difference in SBFL score between AG-tests and MC-tests increased as the nesting depth of the code became deeper, indicating that AG-tests became less effective for fault localization in deeply nested code.

These findings highlight a fundamental difference in the strengths of MC-tests and AG-tests, indicating that each has distinct advantages depending on the structure of the code being tested. While AG-tests are effective in achieving high coverage—particularly for simple conditions and control-flow patterns—they often struggle with fault localization in deeply nested or semantically complex code.

In contrast, MC-tests—crafted with human insight and an understanding of the program’s intent—tend to perform better in fault localization, especially in structurally complex scenarios. These tests are often designed to target edge cases, semantic nuances, or logically challenging paths that automated tools may overlook.

This trend underscores the limitations of automated test generation, which primarily targets structural coverage but may fail to produce semantically meaningful test cases in complex contexts.

These findings suggest that neither MC-tests nor AG-tests alone are sufficient. Rather, an effective testing strategy should adopt a hybrid approach that leverages the strengths of both.
\begin{itemize}
\item AG-tests are well-suited to achieving broad coverage in shallow or syntactically simple code.
\item MC-tests are crucial for fault localization in complex or deeply nested code, where human reasoning can more effectively guide test design.
\item A combined strategy allows testers to balance efficiency (through automation) and diagnostic power (through manual insight), resulting in more reliable and informative test suites.
\end{itemize}

Future work should explore intelligent test generation frameworks that dynamically analyze the structural and semantic characteristics of the code and recommend an optimal mix of MC-tests and AG-tests. Such adaptive strategies could significantly improve both testing efficiency and fault localization accuracy.

\section{Threats to Validity}

In this section, we discuss the potential threats to the validity of our study and describe the measures taken to mitigate them.

\subsection{Internal Validity}
Our results may be influenced by factors such as the selection of EvoSuite configuration options, the parameters used for mutant generation, and the process of test execution. We used the default settings for EvoSuite and Mutanerator to ensure reproducibility; however, different configurations or versions may yield different results. Additionally, the quality of developer-written tests in Defects4J may vary from project to project, potentially affecting the fairness of the comparison.

\subsection{External Validity}
The generalizability of our findings may be limited by the scope of our experimental subjects. We focused on two Java projects (Lang and Math) from the Defects4J dataset, which may not represent all types of real-world software. Furthermore, we considered only Java programs and one automatically test generation tool (EvoSuite). Our conclusions may not directly extend to other programming languages, software domains, or test generation tools.

\subsection{Construct Validity}
The study uses code coverage (statement and branch coverage) and the SBFL score as evaluation metrics. While these are widely used and accepted metrics, they may not capture all aspects of test effectiveness, such as the ability to detect real faults or the maintainability of the test suites. In addition, the SBFL score, while providing insight into fault localization suitability, does not necessarily reflect real-world debugging effort or developer productivity.

\subsection{Conclusion Validity}
We applied statistical tests (Wilcoxon signed-rank test) to compare MC-tests and AG-tests. However, the statistical power of these tests may be limited by the sample size and the characteristics of the datasets used. The possibility of Type I and Type II errors remains, and the conclusions should be interpreted with caution.

Despite these threats, we believe that our study provides useful insights into the comparative strengths and weaknesses of manually and automatically generated test suites. Future work should include a broader range of projects, tools, and evaluation metrics to further strengthen the validity and generalizability of the findings.

\section{Related Work}

Fraser et al. investigated the extent to which multiple test automation tools can help identify fault locations~\cite{fraser}. Their study focused on three tools: Randoop, EvoSuite, and Agitar~\cite{agitar}. The results showed that while each tool individually detected no more than 19.9\% of faults, the combined use of these tools allowed for the detection of 55.7\% of faults. This highlights the complementary nature of different tools in improving fault detection effectiveness.

In line with this idea of complementarity, Serra et al. conducted a comparative study between manually created test cases (MC-tests) and automatically generated test cases~\cite{serra2019msr}. Their evaluation involved tests generated by EvoSuite, Randoop, and JTExpert~\cite{jtexpert}, and was conducted from three perspectives: code coverage, mutation score, and bug detection capability. The results revealed that automatically generated tests performed well in terms of coverage and mutation score, but fell short compared to manually created tests in detecting real bugs. The authors suggested that combining MC-tests and automatically generated tests could help mitigate the limitations of each.

Building on this notion, Roslan et al. proposed an enhanced version of EvoSuite, called EvoSuite\textsubscript{Amp}, which uses developer-written tests as seeds to guide test amplification~\cite{Roslan2022}. In a comparative evaluation with DSpot across 42 versions in the Defects4J dataset, EvoSuite\textsubscript{Amp} achieved higher mutation scores and killed more mutants in many cases. However, the generated tests were often large and suffered from reduced readability, indicating a trade-off between test strength and maintainability.

In addition to improvements in methodology, several studies have investigated the limitations of current test generation tools. Herlim et al. conducted an empirical study on EvoSuite using the SBFT 2020 tool competition benchmark~\cite{Herlim2021}. They analyzed the branches that EvoSuite failed to cover and classified the causes into four categories: object construction issues, object-oriented design constraints, large search spaces, and miscellaneous issues. This classification highlighted the technical challenges that limit EvoSuite's coverage.

Watanabe et al. further examined how the structure of the program under test affects the effectiveness of automatically generated test suites~\cite{watanabe2023profes}. Their study analyzed test suites generated by EvoSuite and identified four main causes of low coverage: specific value requirements, type constraints, unreachable code, and multi-threaded processing. While they proposed mitigation strategies such as inserting dummy branches, they noted that challenges like multi-threaded processing would require improvements in the test generation tools themselves.

More recently, large language models (LLMs) have emerged as a novel approach to test generation. Bhatia et al. explored the use of ChatGPT to generate unit tests for Python programs and compared its performance to that of Pynguin~\cite{Bhatia2024llm4code}~\cite{pynguin}. Their evaluation considered coverage, accuracy, and readability. ChatGPT achieved coverage comparable to or better than Pynguin in some cases but produced incorrect assertions in approximately one-third of the generated tests. Moreover, the uncovered statements differed significantly between the two tools, suggesting that combining them could improve overall test suite completeness.

Finally, some studies have investigated the downstream impact of test suite composition. Matsuda et al. examined how the composition of test cases affects the performance of automated program repair~\cite{Matsuda2020jssst_en}. They manipulated test suites for five types of bug patterns and evaluated the number and correctness of generated patches, as well as the time required for repair. Their findings indicated that adjusting the ratio of passing and failing test cases according to the bug type is critical, and that increasing the number of successful test cases was particularly effective in preventing overfitting during repair.

\section{Conclusion}

In this study, we compared the SBFL score and coverage of manually created tests and automatically generated tests. We used Java projects from the open-source repository Defects4J, employing developer-created test cases as MC-tests and EvoSuite for automated test generation.

The objective of this study was to compare manually created and automatically generated tests from two perspectives: coverage and SBFL  score, in order to identify their respective strengths and weaknesses. Through this comparison, we aimed to gain insights into how MC-tests and AG-tests should be combined and in what scenarios each should be utilized.

This study reveals that MC-tests and AG-tests exhibit different strengths depending on the complexity of the code. While AG-tests are effective at achieving high coverage in simple code, they are less reliable for fault localization when applied to deeply nested structures. In contrast, MC-tests perform better in such complex scenarios, thanks to human reasoning and contextual understanding.

These findings suggest that an effective testing strategy should combine both approaches. AG-tests contribute to broad structural coverage, whereas MC-tests improve diagnostic accuracy. Future work should focus on developing tools that can intelligently suggest the appropriate balance between MC-tests and AG-tests, based on the structural and semantic characteristics of the code.

\begin{credits}
\subsubsection{\ackname}

This research was supported by JSPS KAKENHI Japan (JP24H00692, JP23K24823, JP22K11985)

\end{credits}
%
%
\bibliographystyle{splncs04}
\bibliography{reference}

\end{document}